\documentstyle[seceq,supplement]{ptptex}

\title{
Non-Poissonian level spacing statistics of classically 
integrable quantum systems based on the Berry-Robnik approach
}
\author{
Hironori {\sc Makino}
and Shuichi {\sc Tasaki}$^{*}$}

\inst{
Department of Human and Information Science, 
Tokai University, C-216, 1117 Kitakaname, Hiratsuka-shi, 
Kanagawa 259-1292, Japan
\\
$^*$Department of Applied Physics, Waseda University, 3-4-1 
Okubo, Shinjuku-ku, Tokyo 169-8555, Japan
}

\abst{
Along the line of thoughts of Berry and 
Robnik,\cite{rf1} we investigated the gap distribution 
function of systems with infinitely many independent 
components, and discussed the level-spacing distribution of 
classically integrable quantum systems.  The 
level spacing distribution is classified into three cases:
Case 1: Poissonian if $\bar{\mu}(+\infty)=0$, 
Case 2: Poissonian for large $S$, but possibly not for small 
$S$ if $0<\bar{\mu}(+\infty)< 1$, and Case 3: sub-Poissonian 
if $\bar{\mu}(+\infty)=1$.  Thus, even when the energy levels 
of individual components are statistically independent, 
non-Poisson level spacing distributions are possible.
}

\begin{document}

\maketitle

\makeatletter
\if 0\@prtstyle
\def\asp{0.5em} \def\bsp{0em}
\else
\def\asp{.3em} \def\bsp{0.3em}
\fi \makeatother

An important property of quantum-classical correspondence 
appears in the statistical property of energy levels 
of bounded quantum systems in the semiclassical limit.  
Universal behaviors are found in the statistics 
of {\it unfolded} energy levels at a given interval, 
which are the sequence of numbers uniquely determined by the energy 
levels using the mean level density obtained from the 
Thomas-Fermi rule.  It is widely known that,
for quantum systems whose classical counterparts are integrable,
the distribution of nearest-neighbor level spacing is characterized 
by the Poisson distribution{\cite{rf2}}, while
for quantum systems whose classical counterparts are strongly 
chaotic, the level statistics are well characterized by 
the random matrix theory which gives level-spacing 
distribution obeying the Wigner distribution.{\cite{rf3}}  

Level statistics for the integrable quantum systems has been 
theoretically studied by Berry-Tabor{\cite{rf2}}, 
Sinai{\cite{rf4}}, Molchanov{\cite{rf5}}, Bleher{\cite{rf6}}, 
Connors and Keating{\cite{rf7}}, and Marklof{\cite{rf8}}, 
and have been the subject of many numerical 
investigations.  Still, its mechanism is not well understood, 
the appearance of the Poisson distributions is now widely 
admitted as a universal phenomenon in generic integrable 
quantum systems. 

As suggested, e.g., by Hannay (see the discussion of Ref.1)), 
one possible explanation would be as follows: For an integrable 
system of $f$ degrees-of-freedom, almost every orbit is generically 
confined in each inherent torus, and the whole region in the phase 
space is densely covered by invariant tori as suggested 
by the Liouville-Arnold theorem.{\cite{rf9}}  In other words, 
the phase space of the integrable system consists 
of infinitely many tori which have infinitesimal volumes 
in Liouville measure.  The energy level sequence 
of the whole system is then a superposition of sub-sequences 
which are contributed from those regions.  Therefore, 
if the mean level spacing of each independent subset is 
large, one would expect the Poisson distribution as a 
result of the law of small numbers.{\cite{rf10}}  
This scenario suggested by Hannay is based on the 
theory proposed by Berry and Robnik.{\cite{rf1}}

The Berry-Robnik theory relates the statistics 
of the energy level distribution
to the phase-space geometry by assuming that the sequence of 
the energy spectrum is given by the superposition of statistically 
independent subspectra, which are contributed respectively
from eigenfunctions localized onto the invariant regions
in phase space.  Formation of such independent subspectra is a 
consequence of the condensation of energy eigenfunctions 
on disjoint regions in the classical phase space and of the 
lack of mutual overlap between their eigenfunctions, 
and, thus, can be expected only in the semi-classical 
limit where the Planck constant tends to zero, $\hbar\to 0$.
This mechanism is sometimes referred to as 
{\it{the principle of uniform semi-classical 
condensation of eigenstates}},{\cite{rf11,rf12}} 
which is based on an implicit state by Berry.{\cite{rf13}}  

In this paper, keeping the above mentioned 
scenario in mind, we derive the gap distribution function 
of systems with infinitely many components, 
and discuss the level spacing statistics of integrable 
quantum systems.

In the Berry-Robnik approach,{\cite{rf1} 
the overall level spacing distribution 
is derived as follows: Consider a system 
whose classical phase space is decomposed into 
$N$-disjoint regions.  The Liouville measures 
of these regions are denoted by 
$\rho_i(i=1,2,3,\cdots,N)$ which satisfy 
$\sum_{i=1}^N\rho_i= 1$.  Let $E(S)$ be 
the gap distribution which stands 
for the probability that an interval $(0,S)$ 
contains no level.  $E(S)$ is expressed by 
the level spacing distribution $P(S)$ as,
%
$E(S)=\int_S^\infty d\sigma \int_\sigma^\infty P(x)dx$.
%
When the entire sequence of energy levels is a product of 
statistically independent superposition of $N$ sub-sequences, 
$E(S;N)$ is decomposed into those of 
sub-sequences, $E_i(S;\rho_i)$,
%
\begin{equation}
E(S;N)=\prod_{i=1}^N E_i(S;\rho_i).
\label{eq:1-2}
\end{equation}
%
In terms of the normalized level 
spacing distribution $p_i(S;\rho_i)$ of 
a sub-sequence, $E_i(S;\rho_i)$ is given by 
$E_i(S;\rho_i)= \rho_i\int_S^\infty d\sigma 
\int_\sigma^\infty p_i(x;\rho_i)dx$, 
and $p_i(S;\rho_i)$ is assumed to satisfy{\cite{rf1}}
%
\begin{equation}
\int_0^{\infty} S\cdot p_i(S;\rho_i)dS =\frac{1}{\rho_i}.
\label{eq:1-4}
\end{equation}
%
This equation is satisfactory 
when the Thomas-Fermi rule for individual 
phase space regions still holds.  

Note that the spectral components are not always unfolded 
automatically in general even when the total spectrum is unfolded.  
However, in the sufficient small energy interval, each spectral 
component obeys a same scaling law (see Appendix A of Ref.14)), 
and thus is unfolded automatically by an overall 
unfolding procedure.  Eqs. ({\ref{eq:1-2}}) 
and ({\ref{eq:1-4}}) relate the level statistics in the 
semiclassical limit with the phase-space geometry. 

In most general cases, the level spacing distribution might be 
singular.  In such a case, it is convenient to use 
its cumulative distribution functions:
$
\mu_i(S)=\int_0^S p_i(x;\rho_i)dx.
$

In addition to Eqs ({\ref{eq:1-2}}) 
and ({\ref{eq:1-4}}), we assume the following two conditions: 
\begin{itemize}
\item Assumption (i): The statistical weights 
of independent regions uniformly
vanishes in the limit of infinitely many regions: 
$\max_i \rho_i \to 0\quad\mbox{as}\quad N\to +\infty$.
\item Assumption (ii): The weighted mean of the cumulative 
distribution of energy spacing, 
$\mu(\rho;N)=\sum_{i=1}^N \rho_i\mu_i(\rho)$, 
converges in $N\to +\infty$ to $\bar{\mu}(\rho)$.  The limit 
is uniform on each closed interval: \ $0\le \rho \le S$.
\end{itemize}
Under assumptions (i) and (ii), eqs.({\ref{eq:1-2}}) 
and ({\ref{eq:1-4}}) lead to the overall level spacing 
distribution whose gap distribution function is given by 
the following formula in the limit of $N\to +\infty$,
%
\begin{equation}
E_{\bar{\mu}}(S)=\exp{\left[-\int_0^S
\left(1-\bar{\mu}(\sigma)\right)d\sigma \right]}
\label{eq:1-10},
\end{equation}
%
where the convergence is in the sense of the weak limit. 
When the level spacing distributions 
of individual components are sparse
enough, one may expect $\bar{\mu}=0$ and the gap 
distribution of the whole energy sequence 
is reduced to the Poisson distribution, 
$E_{\bar{\mu}=0}(S)=\exp{\left(-S\right)}$.  
In general, one may expect $\bar{\mu}\not=0$ which corresponds 
to a certain accumulation of the levels of individual components. 

In what follows, starting from 
eqs.({\ref{eq:1-2}}) and ({\ref{eq:1-4}}), 
and assumptions (i) and (ii),  Eq.({\ref{eq:1-10}}) 
is derived in the limit of $N\to +\infty$, and by analyzing 
Eq.({\ref{eq:1-10}}), the level spacing distribution is discussed.

Following the procedure by Mehta,{\cite{rf15}}
we rewrite $E(S;N)$ in terms of the 
cumulative level spacing distribution $\mu_i(S)$ 
of independent components:
\begin{equation}
E(S;N)=
\prod_{i=1}^N
\left[\rho_i
\int_S^{+\infty}d\sigma
(1-\mu_i (\sigma) )\right]
=\prod_{i=1}^N
\left[ 1-\rho_i\int_0^{S}
d\sigma (1-\mu_i (\sigma)) \right].
\end{equation}
The second equality follows from Eq.({\ref{eq:1-4}}), 
integration by parts and 
$\lim_{\sigma\to +\infty}\sigma 
\left(1-\mu_i(\sigma)\right)=0$, which results from the 
existence of the average.  Since the convergence of 
$\sum_{i=1}^N \rho_i \mu_i(\sigma) \to  {\bar \mu}(\sigma)$ 
for $N\to +\infty$ is uniform on each interval 
$\sigma \in [0,S]$ by Assumption (ii), and $|\mu_i(\sigma)|\le 1$, 
$E(S;N)$ has the following limit in $N\to+\infty$:
$$
\log E(S;N)=-\int_0^S d\sigma \left[1-\mu(\sigma;N)\right]
+\sum_i^N O(\rho_i^2)
\longrightarrow - \int_0^S d\sigma 
\left[1-\bar{\mu}(\sigma)\right],\nonumber
$$
where we applied the expansion 
$\log(1+\epsilon)=\epsilon+O(\epsilon^2)$ in 
$\epsilon\ll 1$, and the following 
property obtained from Assumption (i):
$
|\sum_{i=1}^N O(\rho_i^2)|\leq C\cdot
\max_i{\rho_i}\sum_{i=1}^N 
\rho_i =C\cdot\max_i{\rho_i}\to 0
\quad\mbox{as}\quad N\to+\infty
\label{lim:O}
$
with $C$ a positive constant.  Therefore, 
we have Eq.({\ref{eq:1-10}}).
We remark that, when ${\bar \mu}(S)$ is differentiable, 
the asymptotic level spacing 
distribution is described as  
$
P_{\bar{\mu}}(S)= \left[(1-\bar{\mu}(S))^2 
+ \bar{\mu}'(S) \right] \exp{\left[-\int_0^S
\left(1-\bar{\mu}(\sigma)\right)d\sigma \right]}.
$
%

Since $\mu_i(S)$ is monotonically increasing 
and $0 \le \mu_i(S) \le 1$, 
$\bar{\mu}(S)$ has the same properties. 
Then, $1-\bar{\mu}(S)\ge 0$ for any
$S\ge 0$ and one has
$
\frac{1}{S}\int_0^S d\sigma 
(1-\bar{\mu}(\sigma))\longrightarrow 
1-\bar{\mu}(+\infty)
\quad\mbox{as}\quad S\to+\infty.
$
According to this limit, the level spacing distribution 
corresponding to eq.({\ref{eq:1-10}}) is 
classified into the following three cases:
\begin{itemize}
\item
Case 1,~$\bar{\mu}(+\infty)=0$: The 
limiting level spacing distribution 
is the Poisson distribution.  
Note that this condition is 
equivalent to $\bar{\mu}(S)=0$ for 
${}^{\forall} S$ because $\bar{\mu}(S)$ 
is monotonically increasing.
\item
Case 2,~$0<\bar{\mu}(+\infty)<1$: 
For large $S$ values, the limiting 
level spacing distribution is well 
approximated by the Poisson distribution, 
while, for small $S$ values, it may 
deviate from the Poisson distribution.
\item
Case 3,~$\bar{\mu}(+\infty)=1$: 
The limiting level spacing distribution deviates 
from the Poisson distribution for ${}^{\forall}S$, 
and decays as $S\to+\infty$ more slowly than does 
the Poisson distribution.  This case will be referred 
to as a sub-Poisson distribution. 
\end{itemize}
One has Case 1 if the individual 
level spacing distributions are derived from scaled distribution 
functions $f_i$ as $p_i(S;\rho_i)=\rho_i f_i(\rho_i S)$,
where $f_i$ satisfy 
$\int_0^{+\infty} f_i(x) dx = 1$ and 
$\int_0^{+\infty} x f_i(x) dx = 1$, 
and are uniformly bounded by a positive constant 
$D$: $|f_i(S)|\le D$
($1\le i \le N$ and $S\ge 0$).  Indeed, one then has
$$
|\mu(S;N)|
\leq \sum_{i=1}^N \rho_i^2\int_0^S
\left| f_i\left({\rho_i}x\right)\right|dx \nonumber \\
\leq DS\max_i\rho_i\sum_{i=1}^N\rho_i
\longrightarrow 0\equiv {\bar \mu}(S).
$$
In general, one may expect $\bar{\mu}(S)\not=0$ which corresponds 
to the non-Poisson distribution.  Such a case 
is expected when there is strong accumulation of the energy 
levels of individual components which leads to the non-smooth 
cumulative distribution function $\mu_i(S)$.  For a certain 
system class, such accumulation is observable.  
One known example is the
rectangular billiard.{\cite{rf16}}  The level spacing distribution of 
this system shows strong accumulation when the aspect ratio 
of two sides of a billiard wall is close to a rational.  Another 
example is the two-dimensional harmonic oscillator whose 
level spacing distribution is non-smooth for arbitrary 
system parameter.\cite{rf2}  
The final example is studied by Shnirelman, 
Chirikov and Shepelyansky, 
and Frahm and Shepelyansky 
for a certain type of system which contains a 
quasi-degeneracy result from inherent 
symmetry(time reversibility).\cite{rf17,rf18,rf19}  
As is well known, the existence of quasi-degeneracy leads 
to the sharp Shnirelman peak at small level spacings.  
Such phenomena will be discussed in forthcoming 
papers.

In most general cases, the integral in 
equation ({\ref{eq:1-10}}) converges 
in $S\ll+\infty$ and then 
$\lim_{S\to+\infty}E_{\bar{\mu}}(S)\not=0$, 
the limiting gap distribution $E_{\bar{\mu}}(S)$ does 
not work accurately.  In such case, however, its 
differentiation still work accurately 
in $S\to+\infty$ limit ( see Ref.14) ), 
and thus the above classification (Case 1--3) 
holds in general.

In this paper, we investigated the gap distribution 
function of systems with infinitely many independent 
components, and discussed the level-spacing statistics of 
classicaly integrable quantum systems.  
In the semiclassical limit, reflecting infinitely 
fine classical phase space structures, 
individual eigenfunctions are expected to 
be well localized in the phase space and 
contribution independently to 
the level statistics.  Keeping this 
expectation in mind, we considered 
a situation in which the system consists 
of infinitely many components 
and each of them gives an infinitesimal 
contribution.  And by applying the arguments of Mehta, 
and Berry and Robnik, the limiting gap distribution 
was obtained which is described by 
a single monotonically increasing 
function $\bar{\mu}(S)$ of the level 
spacing $S$. Three cases are distinguished: 
Case 1: Poissonian if $\bar{\mu}(+\infty)=0$, 
Case 2: Poissonian 
for large $S$, but possibly not for small 
$S$ if $0<\bar{\mu}(+\infty)< 1$, 
and Case 3: sub-Poissonian if $\bar{\mu}(+\infty)=1$. 
Thus, even when the energy levels of individual components are 
statistically independent, non-Poisson level spacing 
distributions are possible.

The authors would like to thank Professor M. Robnik 
and A. Shudo for their helpful advice.  The authors also thank the 
Yukawa Institute for Theoretical Physics at Kyoto University. 
Discussions during the YITP workshop YITP-W-02-13 on "Quantum chaos: 
Present status of theory and experiment" were useful to 
complete this work.  This work is partly 
supported by a Grant-in-Aid for Scientific Research 
(C) from the Japan Society for the Promotion of Science, and by 
the Ministry of Education, Science, Sports and Culture, 
Grant-in-Aid for Young Scientists (B)(15740244,2003-2005).

\end{document}